# Exploration of lattice Hamiltonians for functional and structural discovery via Gaussian Process-based Exploration-Exploitation

Sergei V. Kalinin,[1,a] Mani Valleti,[2] Rama K. Vasudevan,[1] and Maxim Ziatdinov[1,3]

[1] The Center for Nanophase Materials Science, Oak Ridge National Laboratory, Oak Ridge, TN 37831
[2] The Bredesen Center, University of Tennessee, Knoxville, TN 37996
[3] The Computational Sciences and Engineering Division, Oak Ridge National Laboratory, Oak Ridge, TN 37831, USA

Statistical physics models ranging from simple lattice to complex quantum Hamiltonians are one of the mainstays of modern physics, that have allowed both decades of scientific discovery and provided a universal framework to understand a broad range of phenomena from alloying to frustrated and phase separated materials to quantum systems. Traditionally, exploration of the phase diagrams corresponding to multidimensional parameter spaces of Hamiltonians was performed using a combination of basic physical principles, analytical approximations, and extensive numerical modelling. However, exploration of complex multidimensional parameter spaces is subject to the classic dimensionality problem, and the behaviors of interest concentrated on low dimensional manifolds can remain undiscovered. Here, we demonstrate that combination of exploration and exploration-exploitation with Gaussian process modeling and Bayesian optimization allows effective exploration of the parameter space for lattice Hamiltonians, and effectively maps the regions at which specific macroscopic functionalities or local structures are maximized. We argue that this approach is general and can be further extended well beyond the lattice Hamiltonians to effectively explore parameter space of more complex off-lattice and dynamic models.

---

[a] sergei2@ornl.gov

Materials discovery, design and optimization for specific functionalities is one of the key challenges facing civilization. For centuries, materials discovery has been based largely on an Edisonian approach, exploring large amounts of materials candidates and processing parameters to achieve desired properties. The 20$^{th}$ century has brought about a science-driven materials approach, where knowledge of thermodynamics and kinetic parameters of the process, along with advances in modelling methods have allowed predictive modelling and knowledge-driven exploration of vast materials spaces. In the 21$^{st}$ century, the synergy of machine learning methods capable of exploring vast amounts of information[1] and first principle modelling has paved the way towards rapid acceleration of theoretical discovery. In parallel, correlative machine learning models are progressively being incorporated in experiment-based materials workflows, combining known functional relations, structural and property databases, and computations towards improved materials prediction[2-4].

One of the main challenges in exploration of the materials space is its vast dimensionality. Indeed, the dimensionality of chemical space of small molecules (up to 300 atoms) exceeds $10^{63}$ [Ref. 5]. For crystalline solids on the mesoscopic level, the potential for atomic substitution gives rise to multidimensional phase diagrams comprising single- and multiple phase regions. Given that many properties of interest tend to peak in the vicinity of a specific phase or phase boundary, as exemplified by the morphotropic and relaxor materials[6], electrocatalysis[7], etc., the search for optimal composition or structure becomes an extremely complex problem, all but obviating the grid searches and necessitating more efficient strategies for materials discovery.

In many cases, the useful leads for materials optimization can be obtained with the simplified models that capture the essential physics of materials behavior and allow to incorporate the experimentally observed phenomenological relationships into model parameters. The classical example of this approach is lattice based models, starting with classical Ising[8], Potts[9], and Heisenberg[8] models that describe the vast range of phenomena from allow thermodynamics, magnetism, and surface adatom structures, and extending all the way to complex Hamiltonians describing exotic magnetic orderings in frustrated systems.[10]

It should be noted that the natural question in application of lattice models is the universality class of the Hamiltonian describing specific materials physics, and the values of the corresponding parameters. Traditionally, these were determined based on the combination of the structural data, e.g. from scattering experiments that defined the symmetry of possible interactions, and the macroscopic property measurements. Recently, the approach for extraction of the type and model parameters from observations of microscopic degrees of freedom based on statistical distance minimization was proposed.[11,12] In general, while the relationship between the real materials parameters and model parameters are usually semiquantitative, the models themselves are amenable to exact analytical or numerical solutions, allowing for explorations of trends and emergent behaviors that can further be used for elucidation of experimental relationships or, on more advanced level, be used as a basis for e.g. transfer learning models.

However, whereas lattice Hamiltonian models offer an extremely convenient and powerful approximation for description of many physical and chemical phenomena, exploration of their parameter spaces still represents a considerable challenge. While in simple cases some phases and functional behaviors can be predicted from general physical considerations, in general increase of the dimensionality of the parameter space makes the analysis of phase diagrams and hence comparison to experiment and forward prediction intractable. Recently, the seminal work by Carrasquila and Melko have spurred an intensive effort in machine learning analysis of the Hamiltonians,[13] where information compression methods are used to effectively map phases in parameter space using microstructural states or their derivatives as descriptors, giving rise to a exponential growth of publications.[14-17]

However, the use of grid -based exploration inevitably leads to dimensionality problem, precluding the efficient exploration of multidimensional space. Furthermore, the phases with structural or functional properties of interest could be discovered only by chance, since many unsupervised methods are likely to mix the poorly represented outliers (e.g. states concentrated on a low dimensional manifold) with noise. This naturally leads to the question whether the parameter space of a Hamiltonian can be explored targeting specific functionalities and microstructures. Recently, Vargas-Hernandez et al.[18] demonstrated a Gaussian process regression-based method for exploring phase diagrams which relied on kernel selection based on Bayesian information criteria (BIC). However, the BIC method uses a point estimate and is likely to be unsuccessful for more complex parameter distributions.

Here, we propose the exploration of the phase space of multiparametric Hamiltonians using Gaussian process-based exploration and exploration-exploitation strategies. Using model Ising Hamiltonians with 1- and 2D parameter spaces on 2D square lattice, we demonstrate rapid exploration of the Hamiltonians targeting specific macroscopic thermodynamic parameters or microstructural descriptors such as configurational statistics or preponderant local ordering.

## 1. Model and simulation

As a model system, we have chosen the extended Ising Hamiltonian on the $N^2$ square lattice. For generality, we introduce the model with the Hamiltonian given by Eq. (1):

$$H(\sigma) = - \sum_{<i,j,k,l>} (J_{kl} + \Delta J_{kl}) \sigma_{ij} \sigma_{i+k,j+l} \tag{1}$$

Here $i, j = 1, .. N$ define the lattice site of a square lattice. The indices $k, l$ define the interaction of the spin with its neighbors. Here, we choose $k, l = -2, .., 2$, rendering the model flexible to allow for interactions with the 24 nearest neighbors, i.e. potentially 24 dimensional parameter space. The model Eq.(1) reduces to classical near neighbor Ising model for $J_{10} = J_{-10} = J_{01} = J_{0-1} = J_c$ and other $J_{kl} = 0$. $\Delta J_{kl}$ is a corresponding bond disorder.

The system was explored using classical Monte Carlo method, with the equilibration for $500 \times N^2$ Monte Carlo steps and data acquisition for the next $500 \times N^2$ Monte Carlo steps. For 2D parameter spaces, smaller models were used to avoid large computational times. To avoid emergence metastable states containing large clusters of antiparallel magnetization for large $J$, when the system is kinetically frozen, the initial state was chosen to be uniformly polarized, $\sigma_{ij}(0) = 1$. We used the classical Metropolis method, where spin flip of each lattice site was attempted at every step. The flip is accepted if the energy is reduced; alternatively, the probability of spin flip was calculated using the Boltzmann distribution Eq. (2), where $\beta$ is the inverse temperature and denominator is the partition function.

$$P_\beta(\sigma_i) = \frac{e^{-\beta H(\sigma_i)}}{\sum_j e^{-\beta H(\sigma_j)}} \tag{2}$$

As one set of outputs, we calculate the classical thermodynamics descriptors such as magnetization, susceptibility, and heat capacity. In addition, we introduce the set of functions describing local microstructures. These include the histogram-based descriptors, i.e. distribution of 32 enumerated possible configurations of spins within each 5x5 region. Based on these histograms, we introduce the proximity function defining the deviation of the model microstructure from the reference one

$$G_H = \sum_{l=1}^{32} (h_l - \hat{h}_l)^2 \qquad (3)$$

Where $h_l$ is the statistics of the *l*-th configuration in the model, and $\hat{h}_l$ is the target value. For example, for ferromagnetic (FM) state only one type of neighborhood is possible, whereas for antiferromagnetic states there are two.

Similarly, we introduce the structure similarity function

$$G_s = max\left(\sum_{k,l} corr(\sigma_{ij} t_{i+k,j+l})\right) \qquad (4)$$

defining the difference between the microstructure of interest $t_{i,j}$ and spin configuration $\sigma_{ij}$ in the Ising model. The maximum is taken over all possible shift vectors *k,l* and symmetries of the $t_{i,j}$, allowing for translational invariance. Here, the structure of interest is defined by repetition (tiling) of a target square matrix. We choose the 4x4 representations of the microstructure, and define several special classes corresponding to the FM order, AFM order, and several "non classical" orderings including double AFM 2x2 order, 2x1, 3x1, and 4x2 (stripe) as shown in Fig. 1 (a). The thermodynamic parameters and similarity functions are evaluated by Eq. (3,4) during the data acquisition step. Note that the structure similarity function *de facto* defines the strength of the order parameter field with the configurations defined by tiling matrices.

To illustrate classical grid-based exploration of such a system, shown in Fig. 1 is the evolution of structural order parameters corresponding to the FM, AFM, and 2x1 and 2x2 orderings. Here, the calculations are performed for fixed (reduced) temperature $T = 2.4$, chosen such that the critical value of J corresponding to FM-PM and AFM-PM transition is close to 1 and -1, respectively.

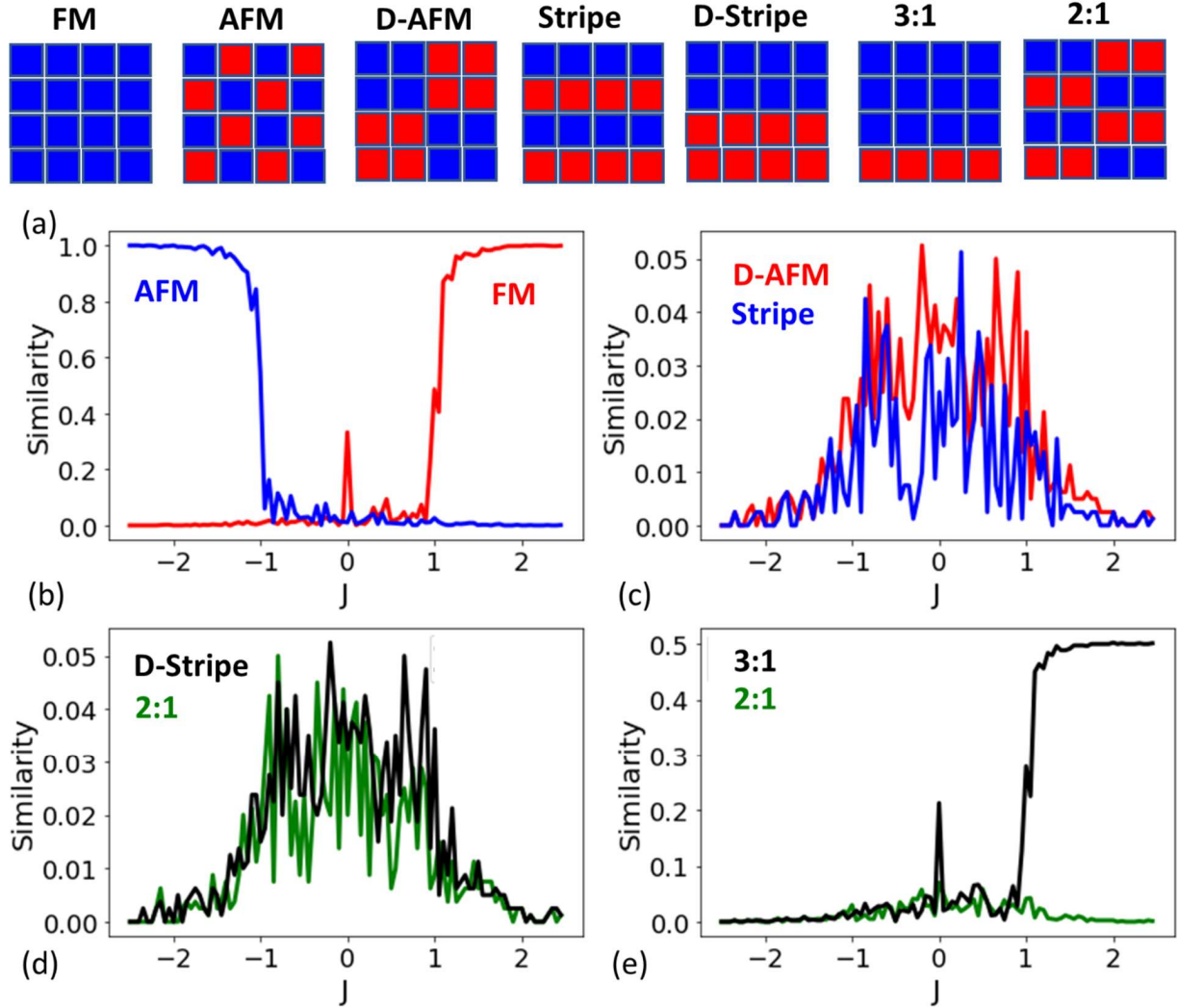

**Figure 1.** Evolution of structural order parameters in the single-parameter Ising model as a function of *J*. Shown are the (a) definitions of the order parameters. The target microstructure is formed by the tiling of the 4x4 elements, and similarity functions allows for translations and 90 rotations. (b) evolution of the classical FM and AFM order parameter, and (c-e) double AFM, stripe, and double stripe order parameters.

The dependence of the order parameters on $J_c$ is shown in Figure 1. Here, Fig.1 (a) shows the definitions of the order parameters. Fig. 1 (b) illustrates the evolution of classical FM and AFM order parameters. Below the AFM phase transition at $J_c \approx -1$ the AFM order parameter is 1 and FM is 0, and the system is in the antiferromagnetic state. Above the FM phase transition at $J_c \approx 1$ the AFM order parameter is 0 and FM is 1, the system is in ferromagnetic states. Finally, for -1 <

$J_c < 1$ both order parameters are zero, the system is in the paramagnetic state. In comparison, shown in Fig. 1 (c) are the order parameters corresponding to the doubled AFM and stripe phases. Both of these are zero in the FM and AFM regions, and non-zero in the PM region. However, the absolute values of these order parameters are small, since they essentially describe the overlap between the randomized configurations of local spins in the paramagnetic phase and the tile defining order parameters. Note that the $J_c$ dependence of these variables can be expected to have non-trivial behaviors when the correlation length and tile become comparable, but the high noise level of these models precludes observation. Similar behavior is observed for other order parameters. Note that non-symmetric (sum of tile matrix elements non-zero) will follow the primary order parameter, but at reduced values. In any case, the value of the similarity function allows us to establish present and possible order parameters, providing a microstructural descriptor for modelling.

A similar approach can be used to explore the mode complex Hamiltonians, as illustrated in Fig. 2 for the next nearest neighbor Hamiltonian. Here, the elements of the $J_{kl}$ matrix are chosen as $J_{kl} = J_c$ for $(k,l) = (0,1), (1,0), (0,-1)$, and $(-1, 0)$ and $J_{kl} = J_s$ for $(k,l) = (1,1), (1,-1), (-1,1)$, and $(-1,-1)$, corresponding to classical next-nearest neighbor (NNN) Ising model. The properties and structure similarity functions are calculated for $T = 2.4$ and both $J_c$ and $J_s$ are chosen to be in the range $(-4, 4)$.

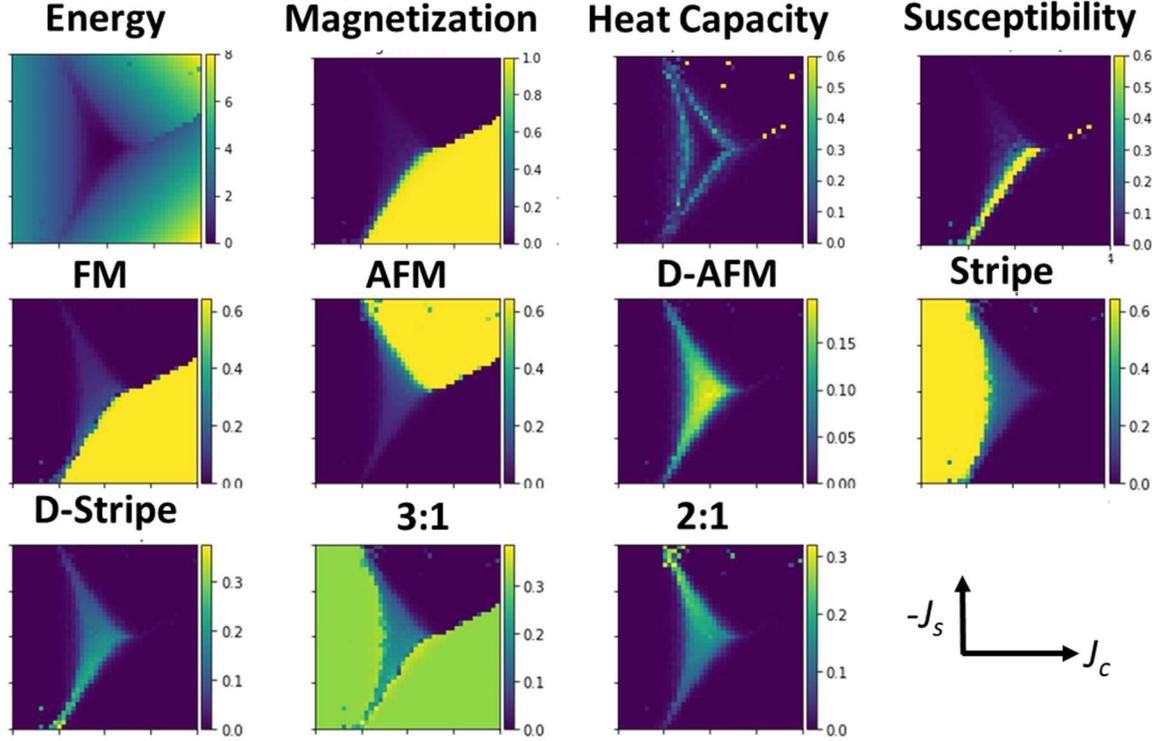

**Figure 2.** The characteristic behavior of 2-parameter Ising model with NN and NNN interactions. Shown are the (top row) classical thermodynamic functions as a function of $J_c$ and $J_s$. Also shown (middle and bottom row) the distribution classical (FM, AFM, stripe) and non-classical OP. The scale for $J_c$ and $J_s$ are (-4, 4).

The characteristic behavior of the NNN Ising model is shown in Fig. 2. The energy per site and the magnetization highlight the ferromagnetic region and the transition between the ferromagnetic, paramagnetic, and antiferromagnetic states. The specific heat and susceptibility delineate the transitions between ferromagnetic, antiferromagnetic, paramagnetic, and frustrated states. Note that whereas heat capacity exhibits anomaly across both the FM-PM and AFM-PM phase transitions, the susceptibility exhibits the anomaly only across the FM-PM transition as expected. Overall, Fig. 2 (top row) provides a traditional approach for mapping the phase diagram of lattice models based on the macroscopic thermodynamic parameter analysis.

The behavior of the order parameters, including existing FM and AFM, and defined double AFM, 2x1 ordering, etc. are shown in Fig. 2. Here, the phase boundaries of the FM, PM, and AFM regions are clearly seen. Similar to the single-parameter Ising Hamiltonian, these are one in the corresponding regions of phase diagram, and close to zero elsewhere. The other *a priori* defined order parameters show roughly constant values in the FM and AFM regions and are zero is the sum of tile elements is zero or finite value if not.

However, exploration of the phase diagrams using grid search is clearly limited. The large overview exploration yields the overall phases, but do not provide the details of critical behavior in the vicinity of phase transition, etc. The scans across the individual phase transitions allow us to obtain a deeper understanding of local physics; however, these require regions of interest to be identified. Secondly, even the overall mapping of phase diagram represents a complex problem if the number of non-zero components of Hamiltonian is 3 or higher. In particular, phases or functionalities concentrated on low-dimensional manifolds in the high-dimensional parameter space can easily be overlooked. This requires developments of efficient methods for exploration of the phase diagram.

## 2. Gaussian process-based exploration

Here, we explore the phase diagram using the Gaussian Process method.[19-22] Gaussian process (GP) regression learns a function $f$ over all source-target pairs $D = \{(x_1, y_1), \ldots (x_N, y_N)\}$, with each pair related by $y = f(x) + \varepsilon$, where $\varepsilon$ is Gaussian observation noise, by performing Bayesian inference in a function space. The key assumption of GP model is that the function $f$ has a prior distribution $f \sim \mathcal{GP}(0, K_f(x, x'))$, where $K_f$ is a covariance function (kernel).[19] The GP process then learn the hyperparameters and sample from the GP to get an estimate of $f$. The kernel function defines the strength and functional form of correlations between the values of the function across the parameter space.

The choice of kernel can be *ad hoc*, based on available functional forms within the chosen GP library and comparing optimization process. The typical example of such functions are the Gaussian radial basis function (RBF), corresponding to weakly informative priors in Bayesian inference. Alternatively, the kernel choice can be guided by the physics of the problem, and the priors and constraints on kernel parameters such as length scale can be obtained based on the domain knowledge about mechanisms of the processes. Note that in all cases, the kernel function is optimized during the GP regression, and kernel parameters such as length scales provide insight into the physics of the process.

GP can be used for approximating continuous non-linear functions from a finite (sparse) number of observations. In these applications, GP is used as an interpolator, where the covariance matrix of the GP posterior distribution serves as an estimator of the uncertainty in the interpolation. As such it can be used for recovering sparse and/or corrupted signal and potentially even enhancing the resolution of multi-dimensional datasets.[23,24] The output of the GP process in this case is the predicted data set, the inferred kernel parameters, and uncertainty maps representing the quality of prediction.

However, the GP can be further used for automated experiment planning. In this, the process starts with presenting the GP interpolator a small number of seed points $\{(x_1, y_1), \ldots, (x_K, y_K)\}$ and initial guesses on kernel hyperparameters (e.g. length scales). Based on these, the GP process optimizes kernel parameters and calculates the predicted and uncertainty maps, similar to interpolation algorithm. The uncertainty can be also used to choose the next point for exploration, e.g. based on maximal uncertainty (exploration) or some combination of uncertainty and function values (exploration-exploitation). Upon calculating the function value (performing the measurement), the set of source-target pairs available to the algorithm and the process is iterated for a given number of iterations, or until the total uncertainty drops below predefined value. This GP is then used for exploration of certain parameter space.

Here we implemented the GP-based exploration based on the Pyro probabilistic programming language[25] and GPim package.[26] Depending on the physics of the problem, we used the RBF or Matern kernels, defined as $k_{RBF}(x_1, x_2) = \sigma^2 \exp\left(-0.5 \times \frac{|x_1-x_2|^2}{l^2}\right)$ and $k_{Matern}(x_1, x_2) = \sigma^2 \exp\left(-\sqrt{5} \times \frac{|x_1-x_2|}{l}\right)\left(1 + \sqrt{5} \times \frac{|x_1-x_2|}{l} + \frac{5}{3} \times \frac{|x_1-x_2|^2}{l^2}\right)$, where $l$ and $\sigma^2$ are kernel length scale and variance, respectively, which are learned from the data by maximizing the log-marginal likelihood.,

Here, the parameter space is defined as dense uniform grid of possible $J$ values. We have chosen the range of (-4, 4) with 100 - 10,000 possible values to allow efficient exploration of system behavior close to phase transitions. Unlike the grid search, these values are not evaluated initially and the grid is used only as discretization of the experimental space. To seed the GP-based exploration process, 3 (for 1 parameter) or 50 (for 2 parameters) values of chosen function, e.g. thermodynamic parameter or similarity to a preselected order parameter, are evaluated in randomly chosen locations.

During the automated exploration of the 1D parameter space model, the learning rate and number of epochs for the GP training are kept constant throughout the process. The numerical values of learning rate are chosen small enough to avoid the unstable exploration, where the algorithm probes the points at the edges of the parameter space, but sufficiently large to allow training in acceptable times. Here, 100 epochs and learning rate of 0.05 were used as default values and tuned if the training was observed not to converge throughout the exploration.

It was further found that the kernel type and parameters should be chosen depending on the function to be explored. Mapping of individual phase regions has shown best results with the use of the RBF kernel with imposed lower bound constraint on the length scales. At the same time, to capture the fine structure of responses such as susceptibility or heat capacity that peak at phase transition, Matern kernel produced best results. The lower bound length scale constraint of kernel function can be adjusted dependent on the sampling in the $J$ space, with the finer grids necessitating lower length scale to avoid "locking" of kernel length scale at grid spacing.

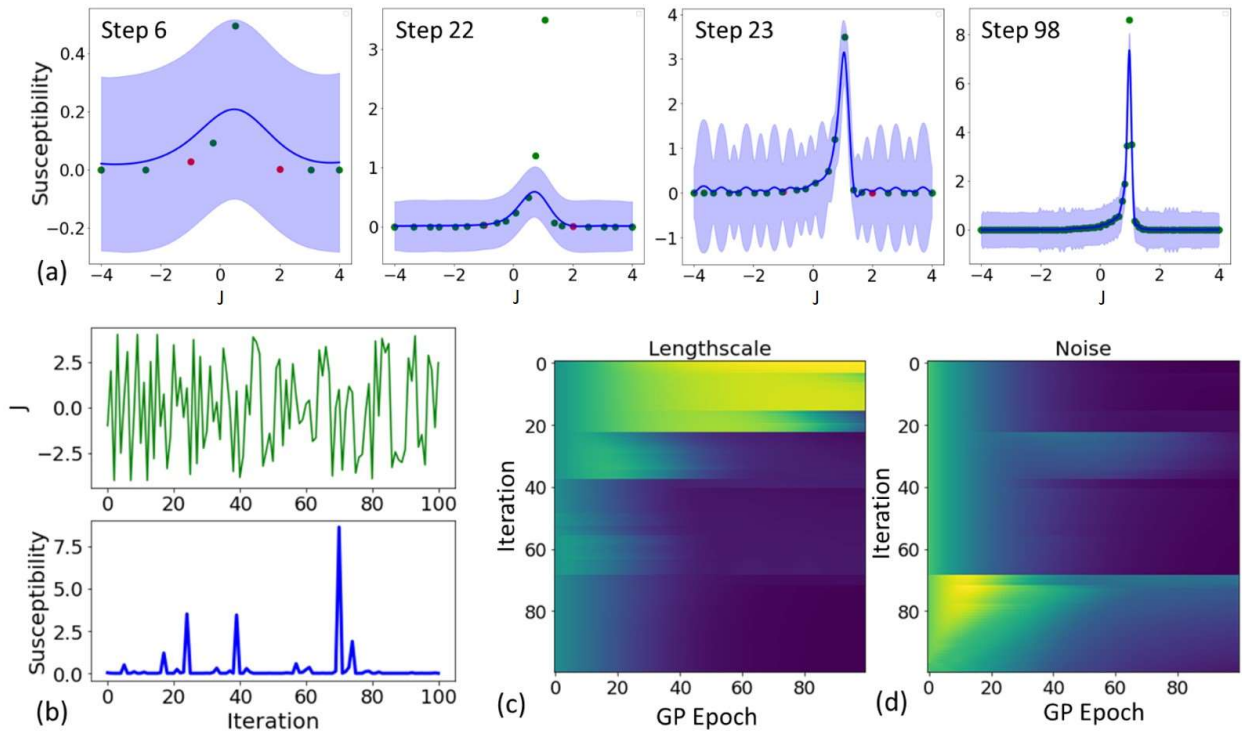

**Figure 3.** Gaussian process exploration of single parameter Ising model using susceptibility as a target function. (a) The GP outputs at several iteration stages. Shown are seed points (red) and points discovered by exploration (green), the GP prediction (blue line) and associated confidence

interval (shaded region). (b) The progression of the exploration process as a function of iteration steps. Shown are the points in the *J* space probed by the process, and the discovered values of susceptibility. Evolution of (c) kernel length scale and (d) noise in GP optimization (horizontal axis) as a function of the number of iterations (vertical axis). Here, lower boundary on kernel length scale was 0.1.

The GP-based exploration of single-parameter Ising model is illustrated in Figure 3. Here, the we explore the 1D parameter space for the behavior of susceptibility. Fig. 3 (a) shows the evolution of the algorithm, including initial seed points, discovered points, predicted mean, and uncertainty. The progression of exploration and the evolution of GP parameters are shown in Fig. 3 (b-d). At the initial stage of exploration process at which points in the proximity to phase transition were not discovered, the GP models the response function as essentially flat, with large kernel length comparable with the interval width, and small uncertainty (top of Fig. 3 (d)). At step 6, the point close to the phase transition retrained the kernel function to a smaller scale, with clearly visible peak close to phase transition. This discovery increased the corresponding uncertainty. On subsequent iterations, the algorithm explored the points with maximal uncertainty both in the vicinity of phase transition and across the whole possible parameter space, looking for additional peaks and readjusting the kernel length scale. This process is further illustrated on transition from step 22 to 23, where discovery of the high response close to the phase transition led to drastic reduction of the kernel length scale. Finally, after 100 iterations, the algorithm converged to efficiently sample the full parameter space.

Notably, the rate of convergence here is fairly slow, since the susceptibility diverges in the vicinity of phase transition, introducing infinitely small length scale in the system. Practically, during the GP-based exploration, the minimal length scale will be limited by the sampling in the *J* space (horizontal) and size *N* of the Ising model (maximal value), etc. As a result, the GP process switches from exploration of *J* space to re-adjustment of the kernel length scale, giving rise to jumps in Fig. 3 (c). Comparatively, exploring parameter space using e.g. magnetization or FM or AFM structural order parameters as input does not lead to such problems, and the process converges rapidly, within 5-10 iterations. The width of the transition however will again be limited by the kernel scale or intrinsic sharpness.

This approach can be further extended to 2D parameter space in Ising system. Here, the exploration space is again formed via dense discretization of $J_c$ and $J_s$ values. It was observed that GP yields large uncertainties at the borders of the exploration space, resulting in the preponderant exploration of these regions in the early stages of the process. To obviate this effect, we additionally introduce a regularization function which, along with its derivatives, is zero at the boundaries of the $J$ space. Here, we have used the $\sin^2$ and $\sin^4$ windows in both directions to avoid this problem. Practically, the exploration behavior depends on the behavior of the product between the tails of the kernel function and regularization function. Similar to the 1D case, some tuning is required for kernel length scales, learning rates, and number of epochs to minimize overall computation time and avoid preponderant exploration of edge regions at the early stages of the process.

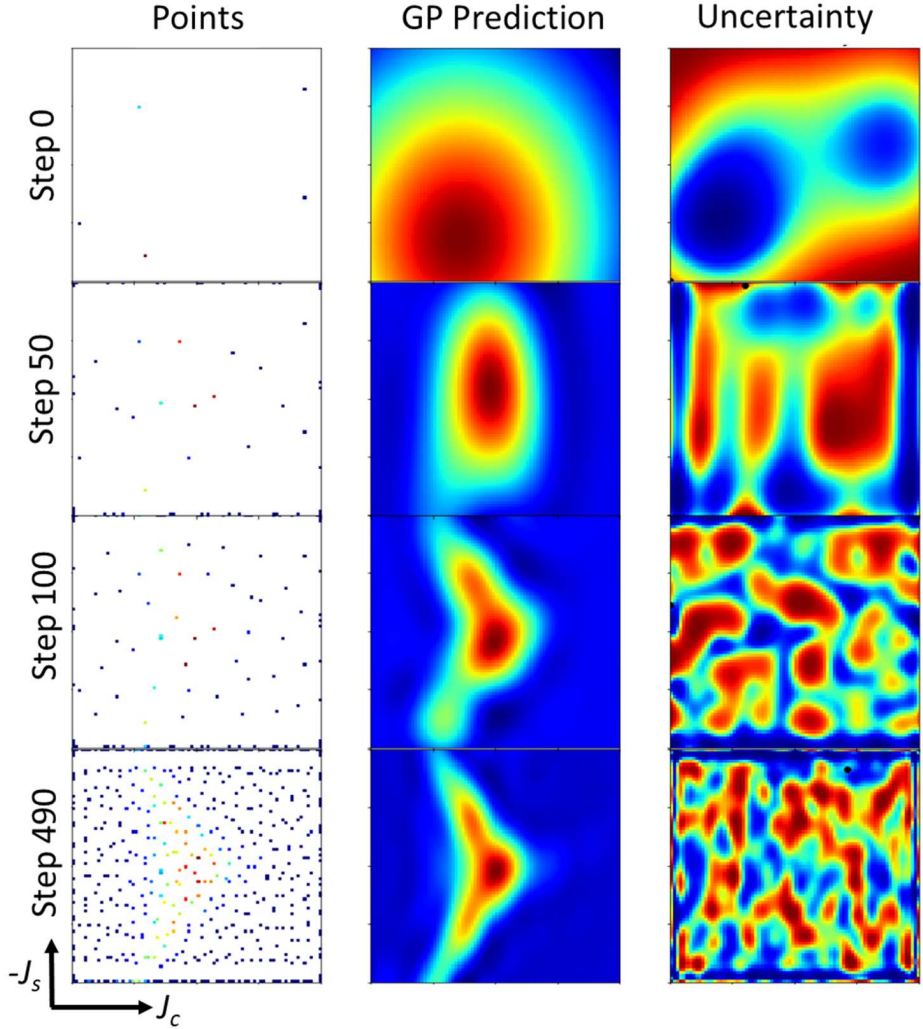

**Figure 4.** Gaussian process-based exploration of Ising Hamiltonian with the 2D parameter space $J_c$, $J_s$ corresponding to nearest- and next nearest neighbor interactions. Shown is the process for RBF kernel, learning rate of 0.05 and lower kernel length of 0.1. The exploration function is the intensity of the double AF phase. The range for both $J_c$ and $J_s$ is (-4, 4). The effect of the window function is clearly seen at Step 490 as a lower density of points close to the edge of exploration space.

Shown in Fig. 4 is the GP-based exploration of the 2D parameter space for nearest- and next nearest neighbor integrals $J_c$ and $J_s$. Here, we have chosen the magnitude of the "non-existent" 2x2 doubled antiferromagnetic phase as a search parameter. The intensity shows clear maximum at the center, corresponding to fluctuations in the paramagnetic phase, and along the lines of the frustrated phases. In this case, the exploration of parameter space is relatively slow, since the phase per se is ill-defined and is concentrated in a relatively narrow region of parameter space. However, even 100 iterations allow to pinpoint this region, whereas 490 given reasonably good localization of the phase region.

Similar behavior can be explored for other thermodynamic and structural parameters. Interestingly, in agreement with the 1D parameter space model, for bulk phases the initial location of the phase regions can be obtained with a much smaller number of iterations, e.g. the localization of the field for the FM phase can be achieved with 50 iterations.

However, the unique aspect of the GP is the opportunity to use it as a basis for Bayesian optimization combining exploration and exploitation. Here, exploration refers to the general minimization of uncertainty on behavior of interest across the parameter space; however, no conditions are imposed for the behavior *per se*. For example, when exploring the heat capacity, the regions with high heat capacity are explored as much as regions with zero heat capacity. However, very often of interest are the regions of parameter space that correspond to maxima of certain behaviors, concentrated in localized and lower-dimensional regions of parameter space. These problems can be addressed by the GP-based Bayesian optimization exploration-exploitation strategies, where an optimization function that guides the selection of the next point comprises both exploration (minimization of uncertainty) and exploitation (maximization of the desired behavior) components.

Here, we implemented the Bayesian optimization strategy consisting of *i*) defining prior and posterior distribution over the objective function *f* using GP, *ii*) using the posterior to derive an acquisition function $\alpha(x)$, *iii*) using the acquisition function to derive the next query point according to $x_{next} = \mathrm{argmin}(\alpha(x))$, *iv*) evaluating *f* in $x_{next}$ and updating the posterior. We note that, prior to our work, the application of Bayesian optimization to a classical Ising model has been studied by Tamura *et al.* with a focus on the computational cost/efficiency.[27] As an acquisition function, we use a simple linear combination of GP-predicted uncertainty and function value. It was found that this strategy offers best results if the GP-based exploration-exploitation process is seeded by a large number of random points in parameter space, and then switched to the exploration-exploitation mode. In principle, a more complex acquisition function (e.g. using non-linear functions of uncertainty and target) and strategy combining switching between exploratory and exploitative stages can be explored (and was found to be beneficial in certain cases), but we defer detailed analysis of these strategies to future studies.

An example of the Bayesian optimization exploration-exploitation of the 1 parameter Ising model targeting the heat capacity is shown in Figure 5. Here, the parameter space was seeded by 20 randomly chosen points. As visible in Fig. 5 (a), several points landed in the vicinity of the heat capacity maxima corresponding to AFM-PM and FM-PM phase transitions. On the exploration-exploitation step, the we see the same dynamics with adjusting the GP kernel and uncertainty; however, once the regions with high response are required, the algorithm varies between the exploration of the parameter space to minimize uncertainty, and concentrate the search next to the maxima. Note that unlike the classical gradient descent methods, the process does not get locked at a single maximum and once one maximum is discovered and explored, the algorithm searches for additional ones.

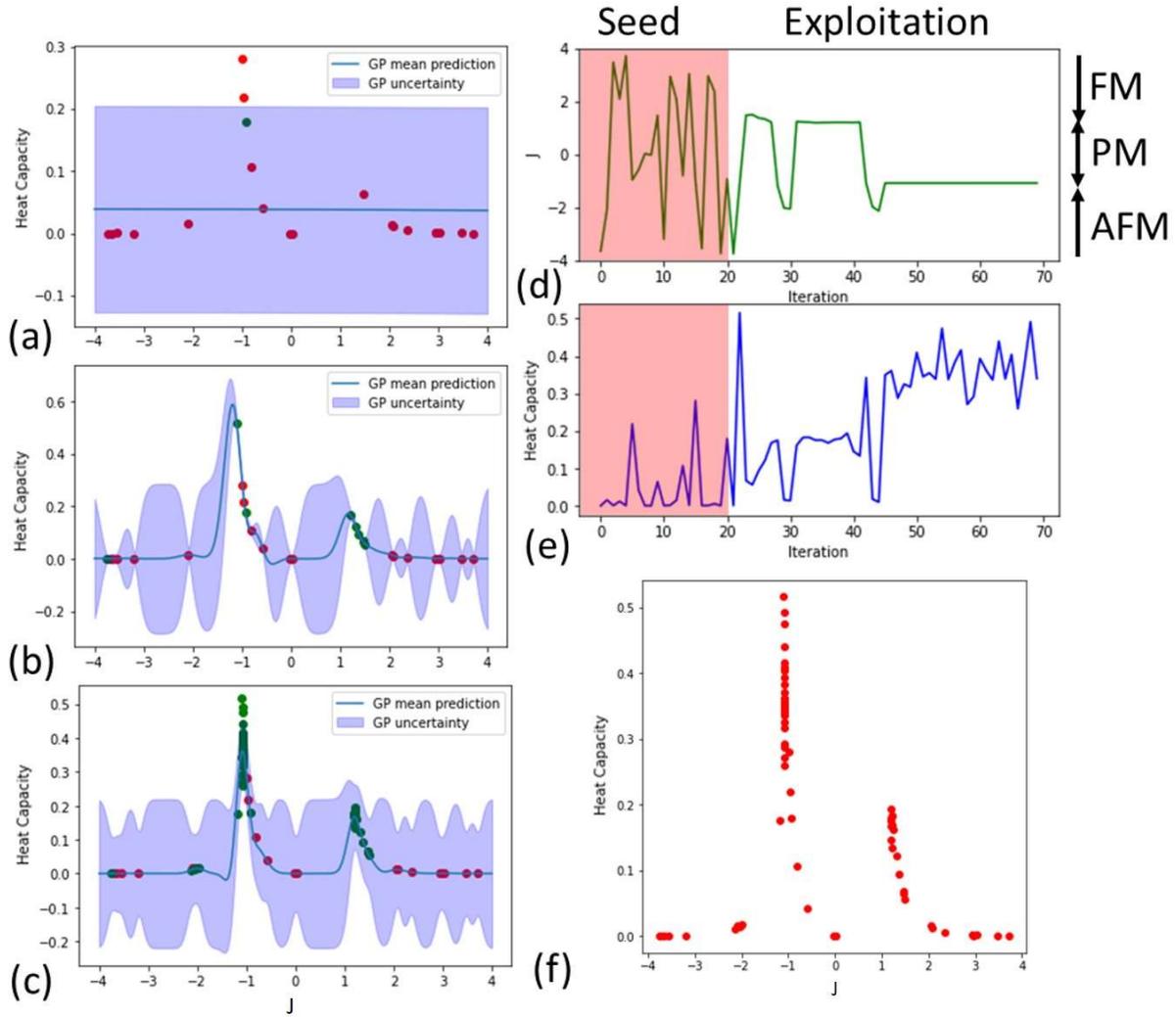

**Figure 5.** GP-based Bayesian optimization for heat capacity in a single-parameter Ising model. (a-c) shows the evolution of the system at the first step (after seeding), and after (b) 8 steps, and (c) 50 steps. (d) Trajectory of exploration-exploitation search after the seeding phase. (e) Discovered values of heat capacity. (f) Final response function. The asymmetry of heat capacities is due to initial conditions (evolution start from magnetized state) and large dispersion due to small size of Ising model and limited number of iterations.

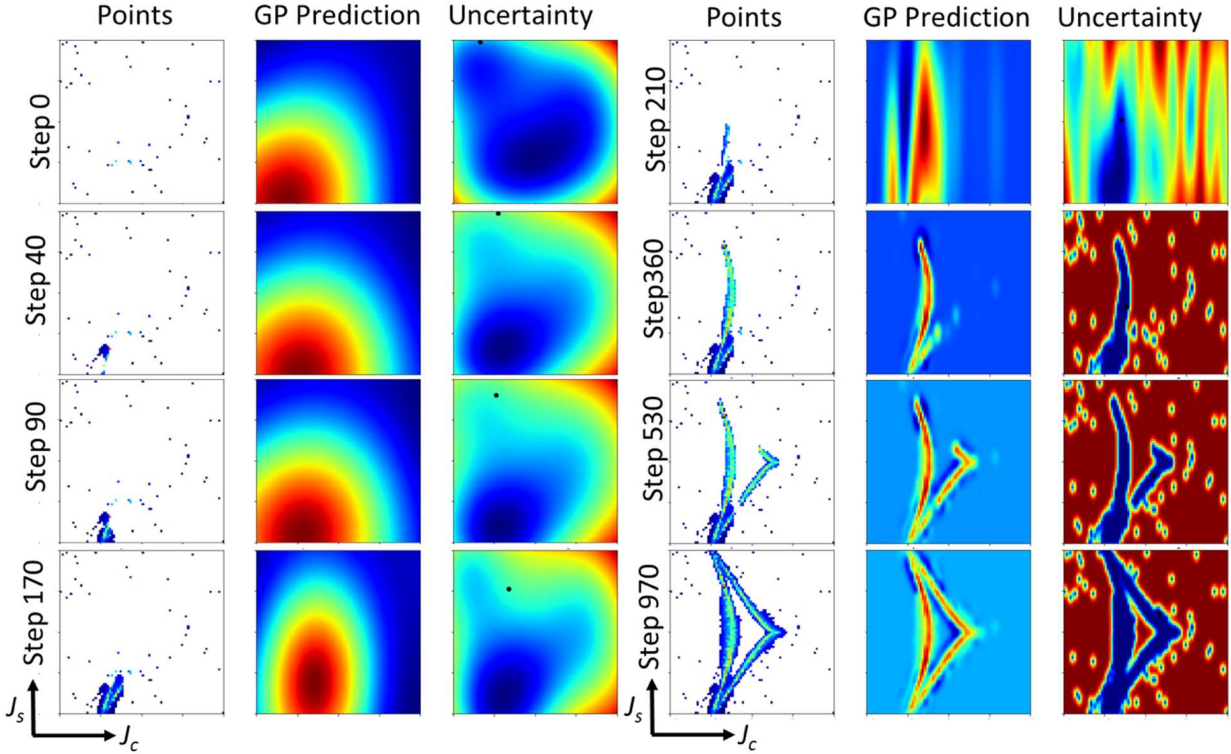

**Figure 6.** GP-based exploration-exploitation of NNN Ising model optimizing heat capacity.

The advantage of exploration-exploitation strategy becomes particularly easy to illustrate for the 2 parameter Ising model, as shown in Figure 6. Here, the process is optimized to minimize the linear combination of GP-predicted uncertainty and function value, aiming to discover the regions with high heat capacity. After the initial seeding stage (stage 0), the process discovers one region in parameter space and seeks to explore adjacent points. Notably, at these early stages the kernel length scale is large, and the process is driven by function maximization. Note that on transition from steps 40 to 90 and 170 to 210, new regions of parameter space are opened, so the algorithm explores values close to the discovered ones. Throughout the process, the algorithm adjusts the kernel function to account for new observations, as can be clearly seen on steps 170, 210, and 360. Finally, from step 360, the kernel length scale converges to the scale (in the J-space) of the feature, and subsequent process explores the regions with high heat capacity. Note that the grid-based representation chosen here implies that once all the regions with sufficiently high heat capacity are samples, the algorithm will switch to exploration on the remainder of parameter space.

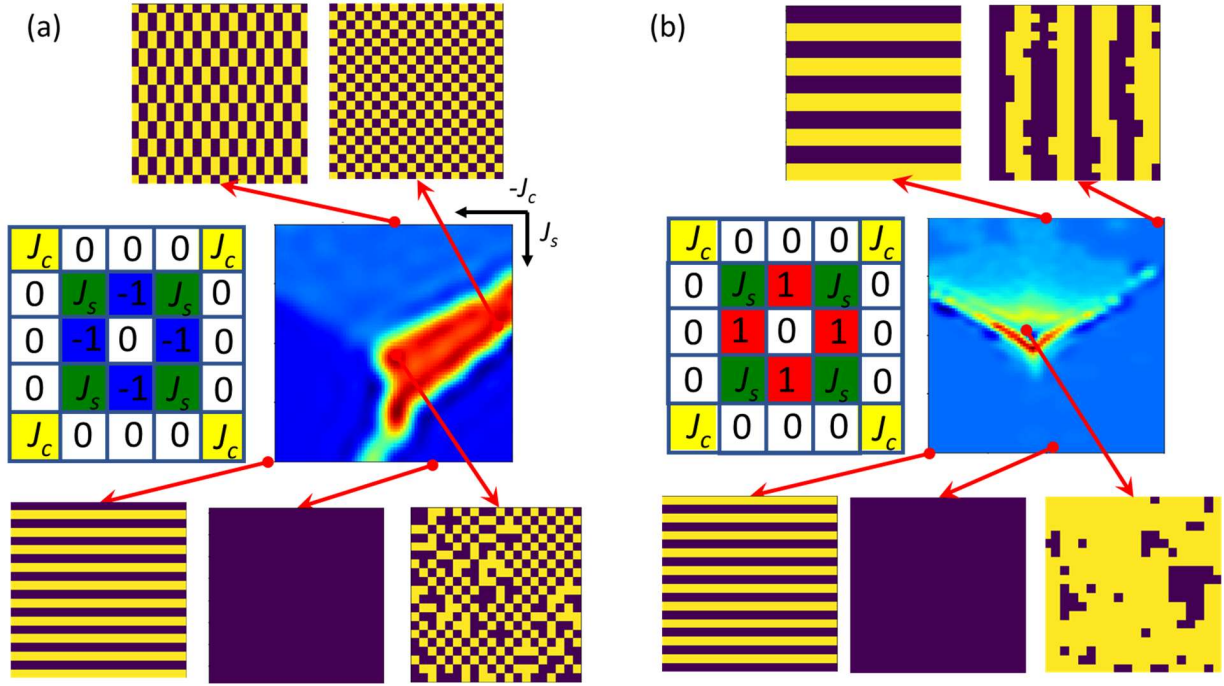

**Figure 7.** Exploration of complex phase diagram for extended (a) antiferromagnetic and (b) ferromagnetic Ising Hamiltonian. Shown is the (a) similarity to antiferromagnetic order parameter and (b) heat capacity. The phase maps are GP mean predictions after 500 exploration steps.

To illustrate application of the GP for exploration of a priori unknown Hamiltonian, shown in Figure 7 are phase diagrams for two ad-hoc chosen extended AFM and FM lattice Hamiltonians. In Fig. 7 (a), shown is the structural phase map for the extended AFM Hamiltonian, complemented by the interactions with the NN and NNN in the (11) directions. Here, GP is used for exploratory search in the structural space, based on the similarity of the structure with the AFM ground state. The resultant phase diagram clearly visualizes the AFM state, stripe-ordered state, 2x1 ordered state, and FM state. Similar analysis is performed for the extended FM Hamiltonian as shown in Fig. 7 (b), but with the heat capacity being visualized. Again, structural phase regions are clearly seen, as well as thermodynamic behaviors at the boundaries. Similarly, the same approach can be trivially extended to the 3D and 4D parameter spaces, but visualization in this case is more complex.

To summarize, here we introduce an approach for rapid exploration of the parameter space of complex Hamiltonians based on specific thermodynamic or structural parameters of interest. This allows rapid mapping of regions of different behaviors both in classical well-studied NN and NNN Ising models, and ad-hoc defined Hamiltonians. While demonstrated here for 1 and 2D systems as limited by the available computation capabilities, we argue that the same strategy can be extended to a higher dimensional parameter space. The strategies for the tuning of the kernel function and regularization are explored. We further note that similar approach is not limited to lattice models and can be used in other modelling methods once some parametrization of parameter space, e.g. force fields in molecular dynamics, etc. is achieved. Here, kernel engineering based on physics of explored system can further accelerate the discovery process.

We further comment on the feasibility for experimental validation of model parameters realization of these systems. The validation of the Hamiltonian model can follow the tried and true method of matching thermal behavior to the models, or recently proposed approach based on the statistical distance minimization between the experimentally observed microstates of the system and the theoretical Hamiltonian system. Similarly, more advanced statistical methods based on the causal machine learning can be used for the analysis of the multimodal experimental data sets.

Experimental tuning of the Hamiltonians, i.e. realization of theoretically discovered phases with functionality or structure of interest represents the obvious challenge. However, presently multiple experimental strategies are available, ranging from classical doping in bulk systems, field effect and substrate effects in the systems such as adatom lattices, and ultimately quantum systems such as trapped ions.

**Acknowledgement**

This work is based upon work supported by the U.S. Department of Energy (DOE), Office of Science, Basic Energy Sciences (BES), Materials Sciences and Engineering Division (S.M.V., R.K.V., and S.V.K.) and was performed and partially supported (MZ) at the Oak Ridge National Laboratory's Center for Nanophase Materials Sciences (CNMS), a U.S. Department of Energy, Office of Science User Facility.

Note: This manuscript has been authored by UT-Battelle, LLC, under Contract No. DE-AC05000R22725 with the U.S. Department of Energy. The United States Government retains and the publisher, by accepting the article for publication, acknowledges that the United States Government retains a non-exclusive, paid-up, irrevocable, world-wide license to publish or reproduce the published form of this manuscript, or allow others to do so, for the United States

**Competing Interests:**

The Authors declare no Competing Financial or Non-Financial Interests

**Data Availability:**

The datasets generated during and analyzed during the current study are available from the corresponding author on reasonable request.

**Author Contribution:**

SVK proposed the concept and led the paper writing. MZ wrote the code for the exploration-exploitation of lattice models with Gaussian processes. RVK and MV wrote the Ising model simulation code. SVK integrated the Ising and Gaussian processes codes, generated and analyzed the data. All the authors contributed to paper writing.